\documentclass[twocolumn,citeautoscript,pra,superscriptaddress,amsmath,amssymb,8pt]{revtex4-1}

\usepackage{graphicx}
\usepackage{color}
\usepackage[dvipsnames]{xcolor}
\usepackage{upgreek}
\usepackage{float}
\usepackage{upgreek}
\usepackage{amsmath,amssymb}
\newcommand{\C}[1]{\ensuremath{^{#1}}C}
\newcommand{\N}[1]{\ensuremath{^{#1}}N}

\newcommand{\ket}[1]{\ensuremath{\left| #1 \right\rangle}}

\begin{document}

\title{Quantum bath control with nuclear spin state selectivity via pulse-adjusted dynamical decoupling}

\author{J. E. Lang}
\email{jacob.lang.14@ucl.ac.uk}
\thanks{These authors contributed equally to this work.}
\affiliation{Department of Physics and Astronomy, University College London, Gower Street, London WC1E 6BT, United Kingdom}

\author{D. A. Broadway}
\thanks{These authors contributed equally to this work.}
\affiliation{School of Physics, University of Melbourne, Parkville, VIC 3010, Australia}
\affiliation{Centre for Quantum Computation and Communication Technology, School of Physics, University of Melbourne, Parkville, VIC 3010, Australia}

\author{G. A. L. White}
\affiliation{School of Physics, University of Melbourne, Parkville, VIC 3010, Australia}
\affiliation{Centre for Quantum Computation and Communication Technology, School of Physics, University of Melbourne, Parkville, VIC 3010, Australia}

\author{L. T. Hall}
\affiliation{School of Physics, University of Melbourne, Parkville, VIC 3010, Australia}

\author{A. Stacey}
\affiliation{Centre for Quantum Computation and Communication Technology, School of Physics, University of Melbourne, Parkville, VIC 3010, Australia}
\affiliation{Melbourne Centre for Nanofabrication, Clayton, VIC 3168, Australia}

\author{L. C. L. Hollenberg}
\affiliation{School of Physics, University of Melbourne, Parkville, VIC 3010, Australia}
\affiliation{Centre for Quantum Computation and Communication Technology, School of Physics, University of Melbourne, Parkville, VIC 3010, Australia}	

\author{T. S. Monteiro}
\email{t.monteiro@ucl.ac.uk}
\affiliation{Department of Physics and Astronomy, University College London, Gower Street, London WC1E 6BT, United Kingdom}

\author{J.-P. Tetienne}
\email{jtetienne@unimelb.edu.au}
\affiliation{School of Physics, University of Melbourne, Parkville, VIC 3010, Australia}

\begin{abstract}

Dynamical decoupling (DD) is a powerful method for controlling arbitrary open quantum systems. In quantum spin control, DD generally involves a sequence of timed spin flips ($\pi$ rotations) arranged to average out or selectively enhance coupling to the environment. Experimentally, errors in the spin flips are inevitably introduced, motivating efforts to optimise error-robust DD. 
Here we invert this paradigm: by introducing particular control ``errors'' in  standard DD, namely a small constant deviation from perfect $\pi$ rotations (pulse adjustments), we show we obtain protocols that retain the advantages of DD while introducing the capabilities of quantum state readout and polarisation transfer. We exploit this nuclear quantum state selectivity on an ensemble of nitrogen-vacancy centres in diamond to efficiently polarise the $^{13}$C quantum bath. The underlying physical mechanism is generic and paves the way to systematic engineering of pulse-adjusted protocols with nuclear state selectivity for quantum control applications.
\end{abstract}

\maketitle 

Quantum baths of nuclear spins typically remain in states close to statistical 50:50 mixtures of spin up and spin down, even for strong magnetic fields, drastically limiting sensitivity and fidelity in many applications ranging from NMR to quantum control using quantum nuclear registers. 
One solution to this challenge is dynamic nuclear polarisation (DNP), the  transfer of spin polarisation from  electron spins  to nuclear spins to hyperpolarise the latter ~\cite{Abragam1978,Bajaj2003}. DNP techniques can be employed to enhance the sensitivity of nuclear magnetic resonance (NMR) detection~\cite{Bajaj2009,Bayro2011} and for initialising nuclear spin-based quantum simulators~\cite{Cai2013}. 

Optically polarised electron spins such as those associated with the nitrogen-vacancy (NV) defect in diamond are particularly interesting for DNP~\cite{London2013,Alvarez2015,King2015,Broadway2018a,Fernandez2018,Shagieva2018,Pagliero2018,Ajoy2018,Schwartz2018} owing to the relatively high ($\sim80\%$) electron spin polarisation achievable on demand, at room temperature~\cite{Doherty2013}. Transfer of NV electron spin polarisation based on tuned cross-relaxation has been used to polarise spins external to the diamond substrate ~\cite{Broadway2018a}. 
Another technique for polarisation transfer is nuclear spin orientation via electron spin locking (NOVEL), which involves continuous driving of the electron spins at a Hartmann-Hahn (HH) resonance with the target nuclei~\cite{Henstra1988,London2013,Fernandez2018,Shagieva2018}. 
Recently, PulsePol, a DD-type protocol allowing polarisation transfer at a rate similar to NOVEL, was proposed~\cite{Schwartz2018}. It is constructed by concatenation of two asymmetric sequences, each made of several electronic spin flips with carefully chosen rotation axes and time delays so as to obtain, on average, an effective flip-flop Hamiltonian for the coupled electron-nucleus spin system. 

Here we propose a radically different approach whereby an asymmetry enabling polarisation transfer is encoded in the spin flip itself, by deliberately introducing a flip-angle adjustment $\delta\theta \neq 0$. That is, instead of  $\pi$ rotations, one deliberately drives rotations of angle $\pi+\delta\theta$. The underlying physical mechanism, explained by Floquet theory, is generic and is found to retain the advantages of DD~\cite{Lloyd1998,Morton2006,Biercuk2009,Du2009,DeLange2010} for decoherence-protected quantum sensing~\cite{Staudacher2013,Shi2015,Lovchinsky2016}. Importantly, it now offers nuclear state selectivity in the sensing as it splits the electron-nucleus resonance into two distinct resonance points, each corresponding to a different quantum nuclear spin state.   

We demonstrate a specific realisation by introducing such a static adjustment in the commonly-used Carr-Purcell-Meiboom-Gill (CPMG) sequence, which consists of a simple train of equally spaced rotations around a fixed axis. The added nuclear state selectivity offers new possibilities to a wide range of single-NV quantum sensing and quantum information experiments including high-fidelity quantum control of weakly coupled nuclear spins. However here we demonstrate its effectiveness in nuclear hyperpolarisation. 
We experimentally implement this protocol, termed PolCPMG, on an ensemble of $\sim 10^5$  NV defects in diamond and demonstrate hyperpolarisation of the surrounding bath of \C{13} nuclear spins as well as real-space imaging of the nuclear polarisation map on a scale of 10's of $\mu$m.

\begin{figure*}
	\includegraphics{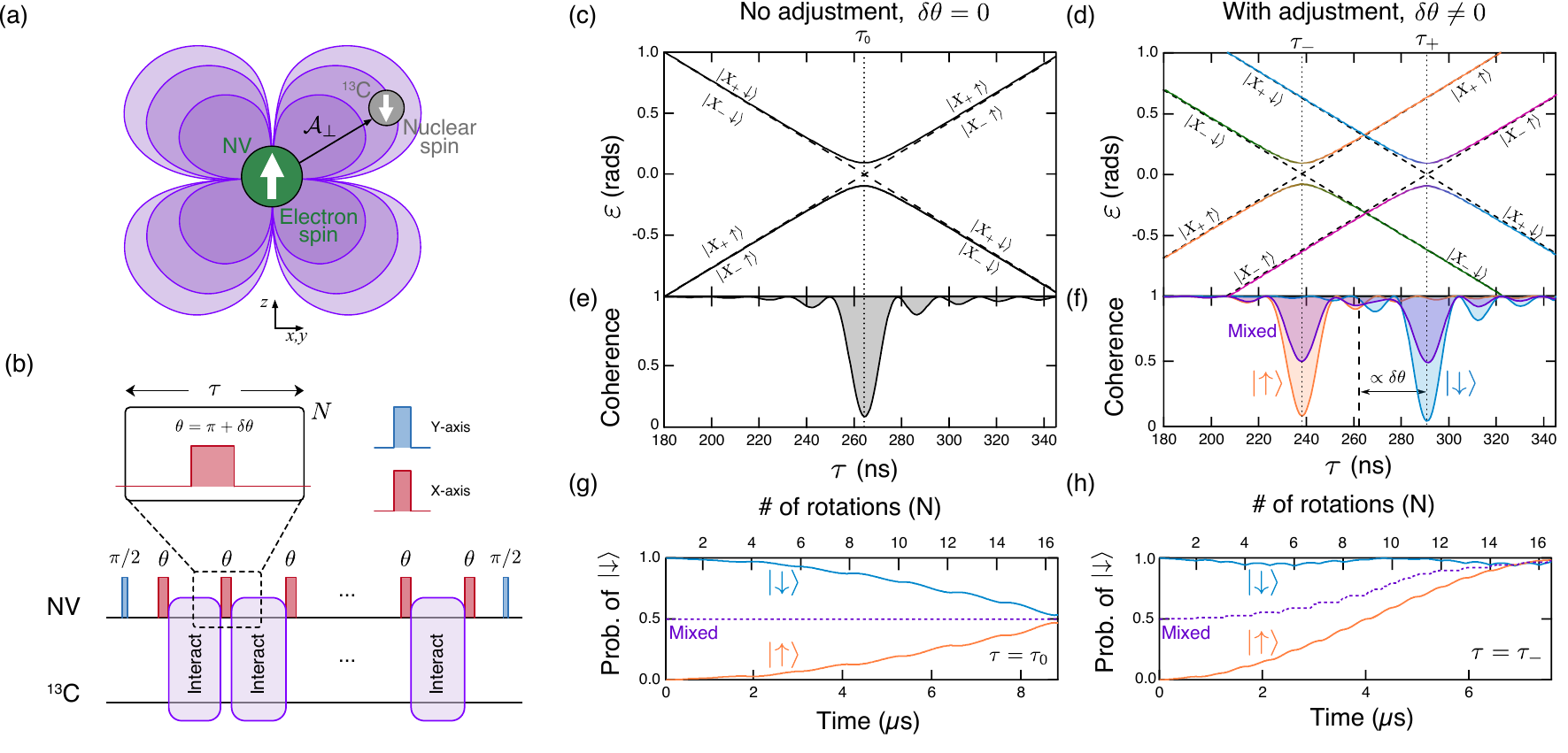}
	\caption{\textbf{Concept: adjusted pulses ($\delta\theta\neq 0$), nuclear state selective resonances and hyperpolarisation.}
		(a) Depiction of a central electron spin (e.g., the NV centre in diamond) surrounded by a bath of nuclear spins (\C{13}). The contour lines represent the transverse hyperfine field, $\mathcal{A}_\perp$, felt by the nuclear spins. 
		(b) Schematic of the PolCPMG dynamical decoupling sequence, which comprises $N$ pulses separated by a period $\tau$. Each pulse rotates the electron spin around the $x$-axis by an angle $\theta=\pi+\delta\theta$, except for the initial and final pulses that rotate the spin by $\pi/2$ around the $y$-axis.
		(c,d) Calculated Floquet phases of the NV-\C{13} coupled system periodically driven by the unit sequence shown in (b), as a function of $\tau$, with $\delta\theta=0$ (c) and $\delta\theta=\pi/10$ (d). Parameters are $\omega_L=1.9$~MHz (corresponding to a magnetic field $B_z=1765$~G) and ${\cal A}_\perp/2\pi=180$~kHz, typical of the experiments described later. Dashed lines correspond to the uncoupled case ($\mathcal{A}_\perp=0$). 
		(e,f) Coherence of the electron spin as a function of $\tau$ after a CPMG (f) and PolCPMG (g) sequence comprising $N=32$ pulses, with the electron spin initialised in \ket{X_+} and the nuclear spin initialised in \ket{\uparrow} (orange), \ket{\downarrow} (blue) or in a completely mixed state (purple). In (e) all the different cases are overlapped and shown in black. 
		(g,h) Time evolution of the nuclear spin state during the CPMG sequence at $\tau=\tau_0$ (g) and during the PolCPMG sequence at $\tau=\tau_-$ (h), with the system initialised as in (f). Here we used ${\cal A}_\perp/2\pi=380$~kHz giving a full polarisation transfer in a relatively small number of pulses ($N=16$), facilitating the visualisation of the pulse-to-pulse evolution.
	}  
	\label{Fig: intro}
\end{figure*}

We consider a system composed of an electron spin $\vec{S}$ coupled with a nuclear spin $\vec{I}$ under a magnetic field aligned along the $z$-axis (Fig.~\ref{Fig: intro}a). The electron spin is subject to a train of microwave pulses with a period $\tau$, with each pulse inducing a rotation of angle $\theta=\pi+\delta\theta$ around the $x$-axis (Fig.~\ref{Fig: intro}b). In practice, the rotation angle may be tuned simply via an appropriate frequency detuning and/or pulse duration. In the frame rotating with the driving microwave field, and neglecting counter-rotating terms, the Hamiltonian reads (see SI)
\begin{equation}
\hat{H}(t) = \omega_L\hat{I}_z + \hat{S}_z\vec{\mathcal{A}}\cdot \vec{I} + \hat{H}_p(t)
\end{equation}
where $\omega_L$ is the nuclear Larmor frequency, $\vec{\mathcal{A}}$ is the hyperfine field felt by the nuclear spin (with a perpendicular projection $A_\perp$ relative to the $z$-axis), and $\hat{H}_p(t)$ is the pulse control Hamiltonian. As the Hamiltonian is periodic, $\hat{H}(t + \tau) = \hat{H}(t)$, Floquet theory provides the natural framework for analysing the dynamics~\cite{Lang2015}. 

The calculated Floquet eigenphases (see details in SI) are plotted as a function of $\tau$ for the case of a standard DD sequence ($\delta\theta=0$, Fig.~\ref{Fig: intro}c) and for the case where a constant flip-angle adjustment $\delta\theta=\pi/10$ is introduced (Fig.~\ref{Fig: intro}d). For the standard  $\delta\theta=0$ case, the  Floquet eigenstates correspond (far from crossings) to the two nuclear spin states, \ket{\uparrow} and \ket{\downarrow}, and are degenerate with respect to the electron spin state. For $\delta\theta \neq 0$, on the other hand,  this degeneracy is lifted to produce four distinct Floquet eigenphases, corresponding to \ket{X_{\pm}}\ket{\uparrow\downarrow} with $\ket{X_{\pm}}=(\ket{0}\pm\ket{1})/\sqrt{2}$, where \ket{0} and \ket{1} are the $\hat{S}_z$ eigenstates of the electron spin. 
In both cases, avoided crossings arise from the presence of a non-zero hyperfine coupling, indicating the periods $\tau$ for which the driven electron spin has, on average, a non-zero interaction with the nuclear spin~\cite{Lang2015}. With $\delta\theta=0$, there is a single (degenerate) avoided crossing at $\tau_0=\frac{\pi}{\omega_L}$. 
This resonance condition is common to many DD sequences including CPMG and XY8, and is routinely used for sensing nuclear spins~\cite{Staudacher2013,Lovchinsky2016}. In a typical NMR sensing experiment, $\tau$ is scanned while monitoring the coherence of the electron spin, producing a spectrum as shown in Fig.~\ref{Fig: intro}e.   

With the flip-angle adjustment (Fig.~\ref{Fig: intro}d), the two avoided crossings occur at two different periods approximately given by (see SI for derivation)
\begin{equation} \label{eq:tau positions}
\tau_\pm\approx\tau_0\left(1\pm\frac{\delta\theta}{\pi}\right).
 \end{equation} 
This leads to two resonances in the coherence spectrum (Fig.~\ref{Fig: intro}f), as recently observed experimentally~\cite{Lang2019}. Importantly, our analysis reveals that these avoided crossings involve pairs of fully orthogonal states for both electron and nuclear spins, for instance the $\tau_-$ crossing mixes the states \ket{X_+}\ket{\uparrow} and \ket{X_-}\ket{\downarrow}. This means that a system initialised in \ket{X_+}\ket{\uparrow} and periodically driven at $\tau=\tau_-$ will undergo oscillations between these two states, as if they were governed by a flip-flop Hamiltonian. Initialisation in \ket{X_+} is naturally done in the CPMG sequence through the initial $\pi/2$ pulse around the $y$-axis when the electron spin is prepared in \ket{0} (Fig.~\ref{Fig: intro}b), which means that a DNP effect is obtained simply by introducing an angle adjustment in the $\pi$ pulses and choosing $\tau$ accordingly. In the following, we will refer to this modified CPMG sequence as PolCPMG.   

To gain more insight into the system's dynamics, we compare the time evolution of the nuclear spin under the standard CPMG sequence at $\tau=\tau_0$ (Fig.~\ref{Fig: intro}g) and under the PolCPMG sequence at $\tau=\tau_-$ (Fig.~\ref{Fig: intro}h). While the \ket{\uparrow} and \ket{\downarrow} states evolve symmetrically under CPMG, resulting in no change of net polarisation for a mixed state, with PolCPMG the \ket{\downarrow} state remains essentially unchanged whereas the \ket{\uparrow} state monotonically evolves to become \ket{\downarrow}. Using Floquet theory, an expression for the total time required to achieve full polarisation transfer 
can be obtained in the limit of instantaneous pulses (see SI),
\begin{equation} \label{eq:transfer time}
t_{\rm PolCPMG} =  \frac{\pi(\pi\pm\delta\theta)}{A_\perp\cos\left(\frac{\delta\theta}{2}\right)}\approx\frac{\pi^2}{A_\perp}
\end{equation}
where the $\pm$ sign corresponds to the $\tau_\pm$ resonances. This is similar to PulsePol~\cite{Schwartz2018} and just a factor $\pi/2$ longer than with NOVEL, under ideal conditions. 

\begin{figure}
	\includegraphics[width=0.9\columnwidth]{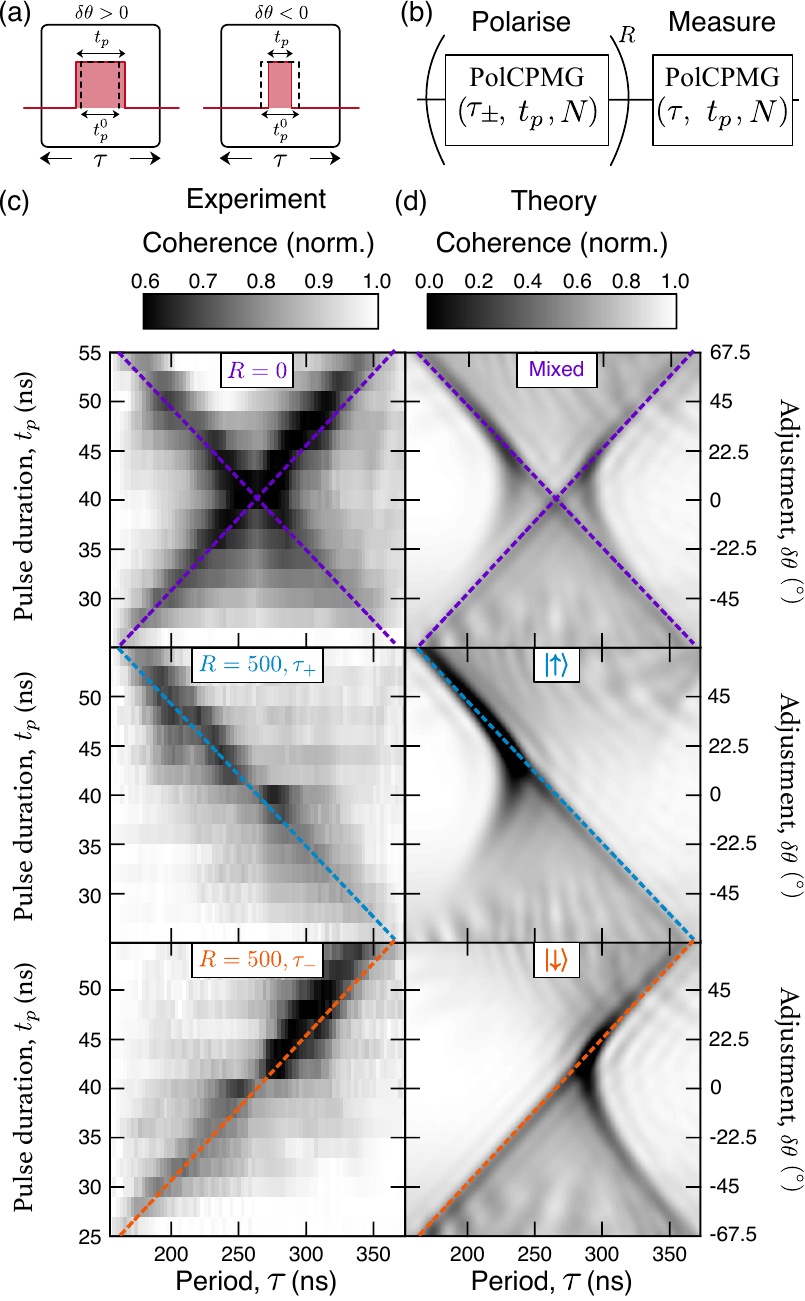}
	\caption{\textbf{Observation of \C{13} hyperpolarisation and nuclear state selective addressing.}
		(a) The flip-angle adjustment $\delta\theta$ is controlled by the pulse duration $t_p$ relative to the nominal duration $t_p^0$ corresponding to a $\pi$ rotation. 
		(b) Sequence used to probe the nuclear polarisation.   
		(c) NV spin coherence (defined as the probability of finding the NV in \ket{0} after the final $\pi/2$ pulse) measured with the sequence shown in (b) as a function of $\tau$ and $t_p$, with $R=0$ (top plot) and with $R=500$ at $\tau=\tau_+$ (middle) or $\tau=\tau_-$ (bottom). Parameters are $\omega_L\approx1.9$~MHz, $t_p^0=40$~ns and $N=32$. 
		(d) Calculated NV spin coherence after a single application of PolCPMG, taking into account the \N{14} hyperfine structure and inhomogeneous broadening (see details in SI). The NV is coupled to a single \C{13} spin (${\cal A}_\perp/2\pi=180$~kHz) initialised in a mixed state (top plot), in \ket{\uparrow} (middle) and in \ket{\downarrow} (bottom).
		In (c,d), the dashed lines correspond to the resonance positions from Eq.~(\ref{eq:tau positions}).
			}
	\label{Fig: spectrum}
\end{figure}

We tested PolCPMG experimentally using the NV centre in diamond as the electron spin, initialised and read out optically, interacting with the bath of \C{13} nuclear spins naturally present in the diamond (1.1\% isotopic abundance). A magnetic field $B_z=1765$~G is applied along the NV axis, giving a \C{13} Larmor frequency $\omega_L\approx1.9$~MHz. The signal from an ensemble of $\sim10^5$ NV centres is measured to average over multiple \C{13} bath configurations. The flip-angle adjustment $\delta\theta$ is controlled via the duration $t_p$ of a rectangular pulse (Fig.~\ref{Fig: spectrum}a). Namely, if $t_p^0=\frac{\pi}{\Omega}=40$~ns is the pulse duration for a $\pi$ rotation, $\Omega$ being the electronic Rabi frequency, then
\begin{equation} \label{eq:delta vs tp}
\delta\theta=\pi\frac{t_p-t_p^0}{t_p^0}.
\end{equation}
We first perform $R$ repetitions of PolCPMG with a fixed period $\tau=\tau_\pm$ to polarise the \C{13} bath, and then probe the state of the bath via a single application of PolCPMG with a variable $\tau$ (Fig.~\ref{Fig: spectrum}b). Coherence spectra obtained for different values of $t_p$, corresponding to an adjustment $\delta\theta$ from $-63^\circ$ to $+63^\circ$, are shown in Fig.~\ref{Fig: spectrum}c for $R=0$ (top) and for $R=500$ at $\tau=\tau_+$ (middle) or $\tau=\tau_-$ (bottom). With no polarisation ($R=0$), the bath is in a mixed state and so two resonances are visible at positions well matched by Eq.~(\ref{eq:tau positions}) (dashed lines in Fig.~\ref{Fig: spectrum}c, top). With the polarisation steps, only one of the two resonances is resolved, indicating that the \C{13} bath has been efficiently polarised in the \ket{\uparrow} and \ket{\downarrow} states for the $\tau_+$ and $\tau_-$ cases (Fig.~\ref{Fig: spectrum}c, middle and bottom), respectively. The variations in amplitude and additional features seen in Fig.~\ref{Fig: spectrum}c originate mainly from the intrinsic hyperfine splitting due to the nitrogen nucleus of the NV (here \N{14}, a spin-1), which means that the NV spin may be driven slightly off-resonance depending on the \N{14} state. We performed numerical simulations of the NV-\C{13} system including the \N{14} hyperfine structure (see details in SI), shown in Fig.~\ref{Fig: spectrum}d. With the \C{13} in a mixed state (top plot), three resonances are resolved near $t_p=t_p^0$, which translates into a single broad line in the experiment. Interestingly, however, the effect of frequency detunings is largely suppressed in certain regimes, especially for $\delta\theta>0$ (as seen by the larger contrast in Figs.~\ref{Fig: spectrum}c,d), which indicates some built-in robustness as discussed later.   

\begin{figure}
	\includegraphics{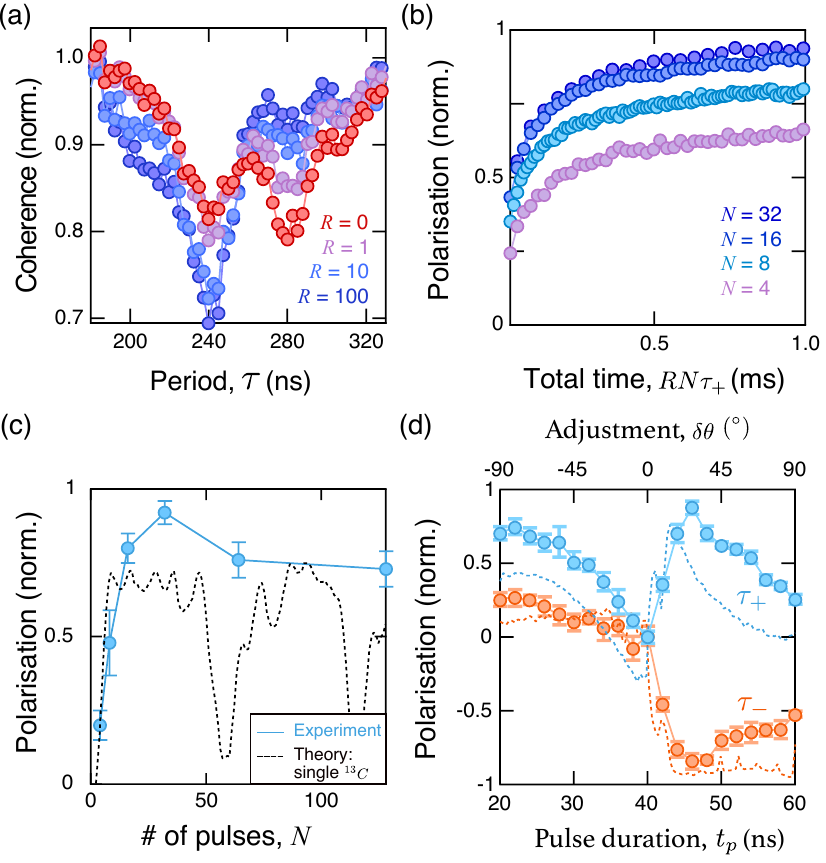}
	\caption{\textbf{Polarisation dynamics.} 
		(a) Coherence spectra taken immediately after $R$ repetitions of the PolCPMG sequence at $\tau=\tau_+=288$~ns, for different values of $R$, with $N=32$. 
		(b) Polarisation of the \C{13} spin bath as a function of $R$ plotted in terms of the total sequence time, $T = RN\tau_+$, for different values of $N$. The polarisation is normalised so that a value of +1 (-1) corresponds to all the \C{13} spins within the NV sensing volume being in the \ket{\uparrow} (\ket{\downarrow}) state.
		(c) Polarisation after $T = 1$~ms as a function of $N$. The dashed line is a numerical simulation for a single \C{13} (${\cal A}_\perp/2\pi=180$~kHz), including the \N{14} hyperfine structure and inhomogeneous broadening. 
		(d) Polarisation after $R=500$ cycles as a function of $t_p$, with $N=32$. For each value of $t_p$, $\tau$ was adjusted to the resulting $\tau_+$ (blue) or $\tau_-$ (orange) resonance. Dashed lines are numerical simulations.
	}
	\label{Fig: Rate}
\end{figure}

To study the dynamics of the polarisation transfer, we vary the number of repetitions $R$ for a given pulse duration, $t_p=44$~ns (i.e. $\delta\theta\approx+18^\circ$), with the period set to $\tau=\tau_+$ polarising the \C{13} bath in the \ket{\uparrow} state. Spectra obtained by scanning $\tau$ immediately after the polarisation reveal a growth of the $\tau_-$ resonance and a suppression of the $\tau_+$ resonance as $R$ is increased (Fig.~\ref{Fig: Rate}a). The relative amplitude of these resonances can be used to estimate the degree of polarisation of the \C{13} spins that are within the sensing volume of the NV probe (corresponding to 5-10 spins typically). This is plotted as a function of the total time $T=RN\tau_+$ in Fig.~\ref{Fig: Rate}b for different numbers of pulses ($N$) per cycle, showing a saturation of the polarisation after a few ms. We find that the polarisation after $T=1$~ms increases with $N$ until $N=32$ before decreasing at larger $N$ (Fig.~\ref{Fig: Rate}c). For a single \C{13}, this optimum would correspond to a coupling ${\cal A}_\perp/2\pi\approx180$~kHz as deduced from Eq.~(\ref{eq:transfer time}), with a minimum expected at $N=64$ (see dashed line), however here this should be averaged over multiple NV-\C{13} coupling strengths. Using $N=32$, we then explored the effect of the choice of $\delta\theta$ (via the choice of $t_p$), varied between $-90^\circ$ and $+90^\circ$. Precisely, for each value of $t_p$, we apply $R=500$ cycles of the PolCPMG sequence at the corresponding $\tau_+$ or $\tau_-$ resonance and then probe the polarisation. The results shown in Fig.~\ref{Fig: Rate}d reveal an asymmetry with respect to $\delta\theta$ whereby positive values of $\delta\theta$ give a stronger polarisation, with an optimum at around $\delta\theta\approx+30^\circ$ for both resonances ($\tau_\pm$). This asymmetry is also present in numerical simulations (dashed lines in Fig.~\ref{Fig: Rate}d) and can be seen in the spectra in Fig.~\ref{Fig: spectrum}d. It originates from the varying degree of robustness of the protocol to a detuning of the microwave driving frequency relative to the NV spin frequency, $\Delta\omega$, for different values of $\delta\theta$, which becomes important when taking into account the \N{14} hyperfine structure and inhomogeneous broadening (see SI).       

\begin{figure*}
	\includegraphics[width=2\columnwidth]{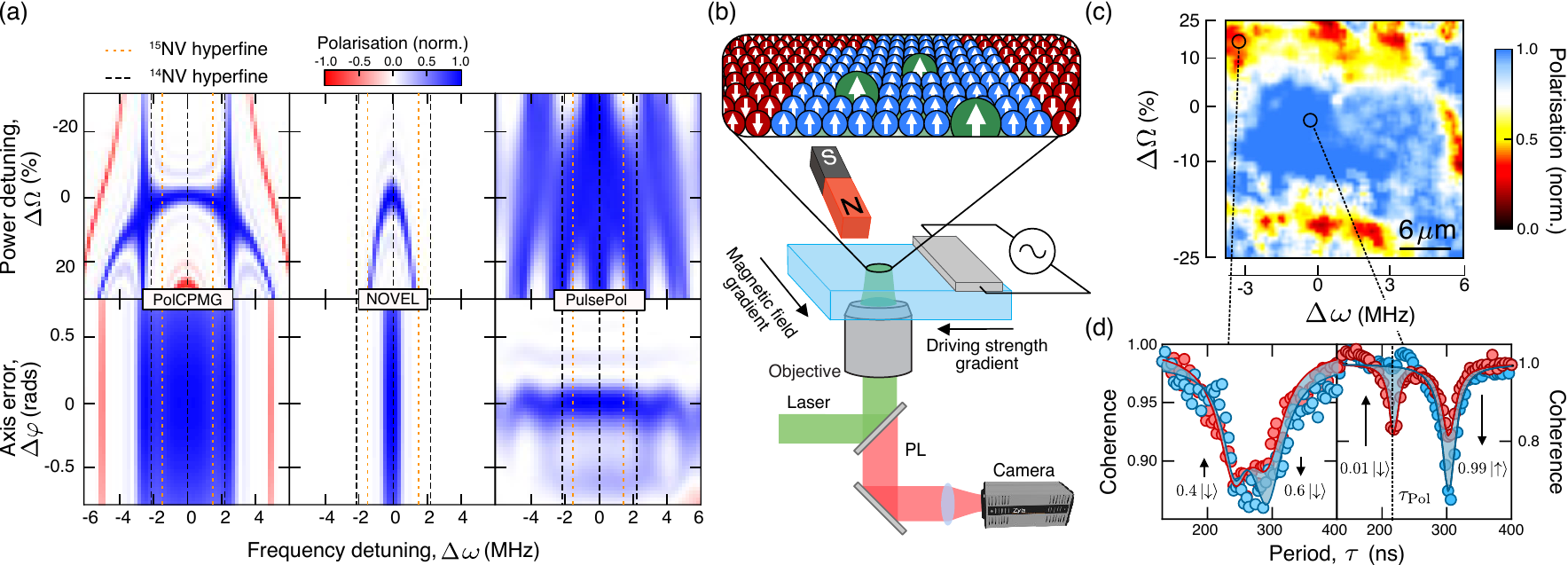}
	\caption{\textbf{Robustness of the protocol.}
		(a) Calculated polarisation of a single \C{13} (${\cal A}_\perp/2\pi=180$~kHz) after applying a single cycle of the PolCPMG (left column), NOVEL (middle) and PulsePol (right) protocols, as a function of $\Delta\omega$ and $\Delta\Omega$ (top colour maps) and $\Delta\varphi$ (bottom). All other parameters are kept constant, set to their nominal values. In all three protocols, the nominal electronic Rabi frequency is $\Omega_0=12.5$~MHz (i.e. $t_p^0=40$~ns). The relative power detuning is defined as $\Delta\Omega=(\Omega-\Omega_0)/\Omega_0$ where $\Omega$ is the microwave driving strength. For PolCPMG, we took $\delta\theta=+30^\circ$, $N=38$ and $\tau=\tau_+=\tau_0(1+\delta\theta/\pi)$. For NOVEL and PulsePol, the total interaction time was adjusted so as to get a perfect polarisation in the absence of detuning. The vertical black (orange) dashed lines indicate the \N{14} (\N{15}) hyperfine transitions of the NV electron spin.
		(b) Experimental setup to image the nuclear polarisation. The coloured arrows depict the NV spins (green), polarised \C{13} (blue) and unpolarised \C{13} (red).
		(c) Polarisation map of a $30\times30~\mu$m region under deliberate gradients of $\Delta\omega$ along $x$ and of $\Delta\Omega$ along $y$. Parameters for the polarisation step are: $t_p=46$~ns, $\tau=\tau_{\rm Pol}=226$~ns, $N=32$, $R = 400$. 
		(d) PolCPMG spectra from two selected regions with low polarisation (left panel) and high polarisation (right panel). The dashed vertical line indicates the value of $\tau=\tau_{\rm Pol}$.
		}
	\label{Fig: Robust}
\end{figure*}

To investigate the robustness further, we calculate the polarisation after one cycle ($R=1$) as a function of $\Delta\omega$, while fixing $\tau$ to the nominal value of $\tau_+$ (determined with $\Delta\omega=0$) at the optimal working point $\delta\theta=+30^\circ$, and compared to the PulsePol and  NOVEL protocols under similar conditions. The resulting polarisation is plotted against $\Delta\omega$ for varying errors in the microwave driving strength, $\Delta\Omega$ (Fig.~\ref{Fig: Robust}a, top plots), and for a range of angle errors between the $x$ and $y$ rotation axes, $\Delta\varphi$ (bottom). We find that at $\Delta\Omega=0$ PolCPMG is nearly as robust to frequency detunings as PulsePol, and significantly more robust than NOVEL. Importantly, the polarisation with PolCPMG remains larger than 80\% for frequency detunings $|\Delta\omega|\lesssim2.8$~MHz, which covers the intrinsic hyperfine structure of the NV spin (\N{14} or \N{15}) as shown by the vertical dashed lines in Fig.~\ref{Fig: Robust}a. 
As for the other possible imperfections, PolCPMG is found to be less robust against $\Delta\Omega$ errors but more robust against $\Delta\varphi$ errors, compared to PulsePol.    

Finally, we use PolCPMG to demonstrate real-space mapping of the nuclear polarisation. Precisely, we use a wide-field imaging setup to illuminate a 1-$\mu$m-thick layer of NV centres near the diamond surface (Fig.~\ref{Fig: Robust}b) and map the polarisation of the \C{13} bath following application of PolCPMG for $T=1$~ms. To test the robustness of the protocol, gradients of $\Delta\omega$ and of $\Delta\Omega$ were deliberately introduced along the $x$ and $y$ spatial direction, respectively, through inhomogeneous applied fields (see schematic in Fig.~\ref{Fig: Robust}b). A polarisation image of a $30\times30~\mu$m$^2$ region is shown in Fig.~\ref{Fig: Robust}c, revealing a polarisation in excess of 80\% in the majority of the image despite a variation of $\Delta\omega=-3.5$ to 6~MHz and of $\Delta\Omega=\pm25\%$. As shown in Fig.~\ref{Fig: Robust}d, the level of polarisation can still be inferred from the PolCPMG spectra even in far-detuned conditions. 

PolCPMG has advantages over existing methods: it is more robust than NOVEL or cross-relaxation and simpler to implement than PulsePol in that it is compatible with standard digital modulation hardware -- whereas PulsePol requires more advanced phase control. The robustness of PolCPMG could be further improved by adapting existing methods of pulse shaping or composite pulses~\cite{Casanova2015} to provide the extra rotation, and there is scope for polarisation speed up through optimisation of the Floquet modes at the avoided crossing. Thus, PolCPMG offers a promising route towards the long-standing goal of large-scale NV-based hyperpolarisation of external samples (realised recently on single NVs~\cite{Broadway2018a,Fernandez2018,Shagieva2018}), which could enable ultra-sensitive NMR spectroscopy for in-line chemical analysis or cell biology studies~\cite{Bucher2018,Smits2019} or form the basis of a quantum simulator~\cite{Cai2013}. In these endeavours, the ability to directly image the nuclear polarisation over 10's of $\mu$m via near-surface NVs as demonstrated here may become an ubiquitous tool. More generally, our work paves the way to systematic engineering of pulse adjustment-enhanced protocols for quantum information processing and quantum sensing.   

\section*{Author contributions}

J.E.L., T.S.M. and J.-P.T. planned and designed the study, from an original proposal by J.E.L. J.E.L. and T.S.M. developed the theory including Floquet model and performed the bulk of the simulations, with additional contributions from G.A.L.W. D.B. and J.-P.T. performed the experiments and analysed the data, with inputs from J.E.L., T.S.M., L.T.H. and L.C.L.H. A.S. synthesised the diamond sample. J.E.L., D.A.B., T.S.M. and J.-P.T. wrote the manuscript with inputs from all authors.

\section*{Acknowledgements}

We acknowledge support from the Australian Research Council (ARC) through grants DE170100129, CE170100012 and FL130100119. This work was performed in part at the Melbourne Centre for Nanofabrication (MCN) in the Victorian Node of the Australian National Fabrication Facility (ANFF). J.E.L. is funded by an EPSRC Doctoral Prize Fellowship. D.A.B. is supported by an Australian Government Research Training Program Scholarship. \\

\bibliography{bib}


\begin{widetext}

\section*{Supporting Information}

\section{Theoretical description of PolCPMG}

\subsection{Rotation adjustments with rectangular pulses}

The Hamiltonian describing an electronic spin qubit subject to resonant microwave control is given by 
\begin{equation}
\hat{H}'_p(t) = \omega_0\hat{S}_z + 2\Omega(t)\hat{S}_x\cos(\omega t + \phi)
\end{equation}
where $\omega_0$ is the frequency splitting of the qubit, $\Omega(t)$ is the microwave drive power which is turned on and off to produce sharp pulses, $\omega$ is the microwave drive frequency and $\phi$ is the pulse phase which sets the desired rotation axis for control pulses. The electronic spin-1/2 operators are given by the usual Pauli matrices $\hat{S}_{x,y,z} = \frac{1}{2}\hat{\sigma}_{x,y,z}$. In the case of NV based systems we assume two levels are selected from the spin-1 system as the qubit. 

In the frame rotating with the microwave frequency (and neglecting fast rotating terms) the Hamiltonian can be re-written as
\begin{equation}
\hat{H}_p(t) = \Delta\omega\hat{S}_z + \Omega(t)\hat{S}_\phi
\label{eq: Pulse Ham}
\end{equation}
where $\Delta\omega = \omega_0 - \omega$ is the frequency detuning of the microwave pulses.

For top-hat pulses, $\Omega(t) = \Omega$ during pulses and $\Omega(t) = 0$ elsewhere (Fig.~\ref{fig: PolCPMG schematic}). In the absence of detuning the pulse length is then set as $t_p^0 = \pi/\Omega$ to achieve a $\pi$-pulse. In reality a combination of pulse duration $t_p = t_p^0 + \delta t_p$ and frequency detuning errors can produce additional small rotations, $\delta\theta$. In PolCPMG the additional rotation is input by design. In the case of no frequency detuning ($\Delta\omega=0$), $\delta\theta$ is related to the pulse duration via 
\begin{equation} \label{eq:deltatheta}
\delta\theta = \Omega\delta t_p = \pi \frac{t_p - t_p^0}{t_p^0}~,
\end{equation}
which is Eq.~(4) of the main text. 

\begin{figure*}[h!]
\includegraphics[scale = 0.6]{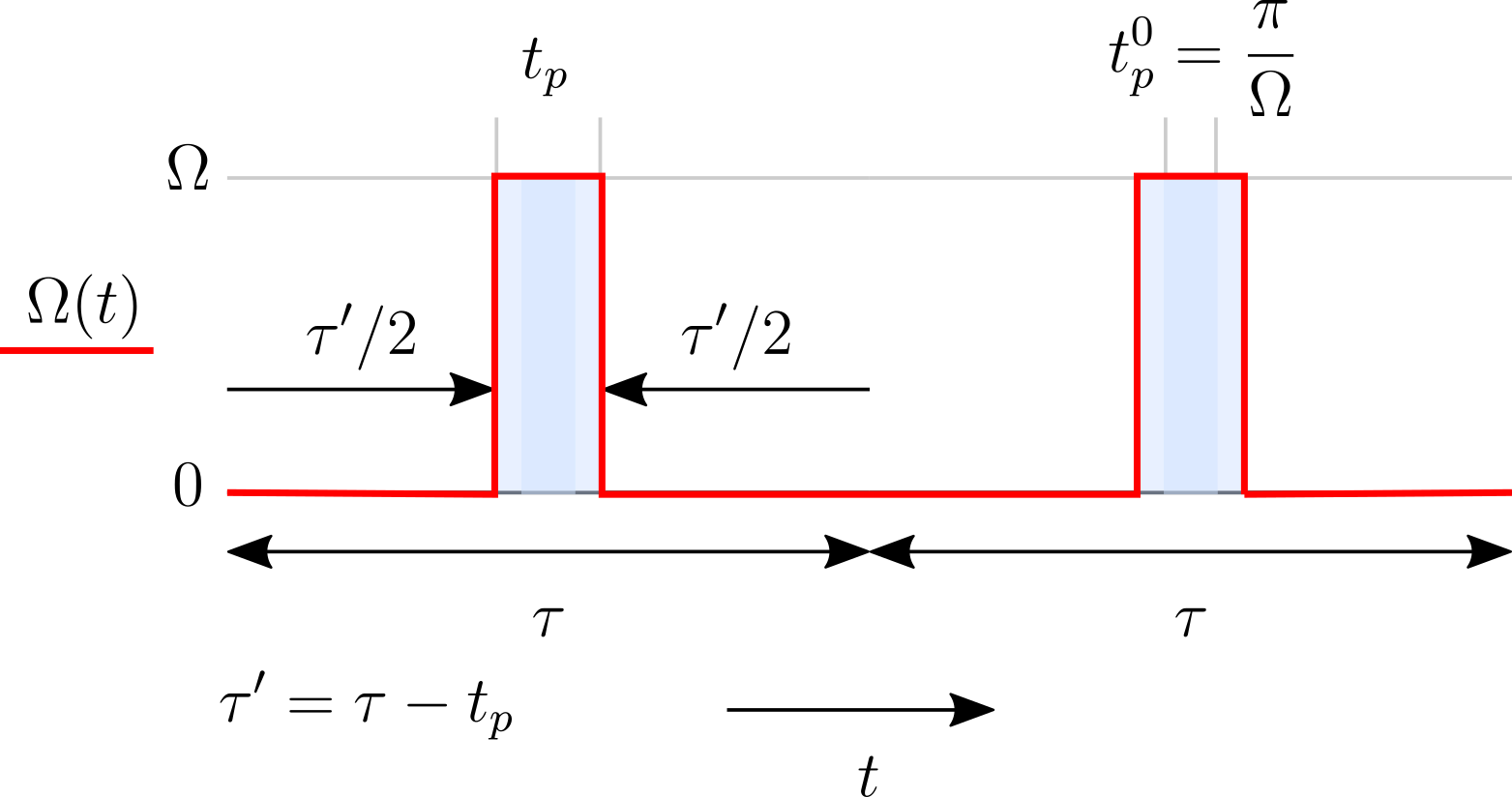}
\caption{\textbf{PolCPMG.} Schematic showing the microwave drive strength, $\Omega(t)$, for the PolCPMG protocol using top-hat pulses. For $t_p = \pi/\Omega$ the standard CPMG sequence is recovered. To achieve nuclear polarisation the sequence is applied repeatedly at $\tau = \tau_\pm$ (defined in the text). All pulses are applied about the $x$-axis.}  
\label{fig: PolCPMG schematic}
\end{figure*}

\subsection{System Hamiltonian and Floquet theory}

For the polarisation of nuclear spins via PolCPMG we study the Hamiltonian
\begin{equation}
\hat{H}(t) = \omega_L\hat{I}_z + \hat{S}_z\vec{\mathcal{A}}\cdot\vec{I} + \hat{H}_p(t)
\label{eq: Ham}
\end{equation}
where $\omega_L$ is the nuclear Larmor frequency, $\vec{\mathcal{A}}\cdot\vec{I} = A_\perp\hat{I}_x + A_\parallel\hat{I}_z$ is the hyperfine coupling and $\hat{H}_p(t)$ is given in Eq.~\eqref{eq: Pulse Ham}. The spin-1/2 nuclear spin operators are given by the usual Pauli matrices $\hat{I}_{x,y,z} = \frac{1}{2}\hat{\sigma}_{x,y,z}$. The usual pure-dephasing approximation has been made to neglect the $\hat{S}_{x,y}$ coupling terms. 

For typical DD control using the CPMG sequence, microwave $\pi$-pulses are applied along the $x$-axis with a regular spacing, $\tau$.  For standard CPMG based sensing, a coherence dip appears at characteristic pulse spacing $\tau = \tau_0 \equiv \pi/|\omega_\text{L}|$ but when an additional rotation is included in the control pulses this dip splits into a pair at $\tau = \tau_\pm$ \cite{Lang2019}. PolCPMG applies $(\pi + \delta\theta)$-pulses resonantly with $\tau_+$ or $\tau_-$ which causes polarisation to transfer from the electronic to nuclear spin -- allowing for the initialisation of nuclear spins. 

Due to the periodicity of the Hamiltonian, Floquet theory can be used to obtain the resonance positions and polarisation transfer rate. Studying the Floquet spectrum -- specifically the position of avoided crossings and their widths -- reveals the system dynamics \cite{Lang2015, lang2017enhanced}.

The Floquet spectrum is given by the eigenphases, $\varepsilon$ (called Floquet phases), of the one-period evolution operator, $\hat{U}(T = 2\tau)\ket{\phi_F} =\exp(-i\varepsilon)\ket{\phi_F}$, where here $\ket{\phi_F}$ is a Floquet mode -- an eigenstate of the stroboscopic evolution. Figure~\ref{Fig: Th_with_finite_tp} shows numerical simulations of example Floquet spectra (a-c) and the electronic response (d-f) for (a,d) standard CPMG detection (equivalent to PolCPMG for $\delta\theta = 0$),  (b,e) a PolCPMG sequence and (c,f) the PulsePol sequence~\cite{Schwartz2018}. Panels (a) and (b) recreate a plot from the main text but here $t_p = 40$~ns (rather than $t_p \rightarrow 0$) which results in the slight asymmetry in avoided crossing widths and coherence dip depths. 

The comparison of Figs.~\ref{Fig: Th_with_finite_tp}b and \ref{Fig: Th_with_finite_tp}c highlights the fundamental difference between PolCPMG and PulsePol. With PulsePol, there is one avoided crossing and one true crossing at a single resonance condition, $\tau=3\pi/\omega_L$, which implies that for a given state of the electronic spin (say \ket{0}), there is an interaction only for a specific state of the nuclear spin (\ket{\downarrow}). This is the reason for the polarisation transfer effect. In contrast, for PolCPMG there are two avoided crossings as in most DD sequences, but these occur at two different periods $\tau_\pm$ and this leads to a polarisation transfer when driving the system at one such period.      

In PolCPMG, the resonance positions, $\tau_\pm$, are set by the positions of the avoided crossings. The positions of the avoided crossings can be found by looking for the \textit{true crossings} in the \emph{unperturbed Floquet spectrum} which are given by the eigenphases of the one-period evolution operator when we set $\vec{\mathcal{A}} = 0$ (dashed lines in Fig.~\ref{Fig: Th_with_finite_tp}(a,b)). This derivation of $\tau_\pm$ is detailed in Section~\ref{sec: res pos}. In Section~\ref{sec: pol rate} the hyperfine coupling is reintroduced ($\vec{\mathcal{A}} \neq 0$) to determine the PolCPMG nuclear polarisation rate. 

\begin{figure*}[h!]
	\includegraphics{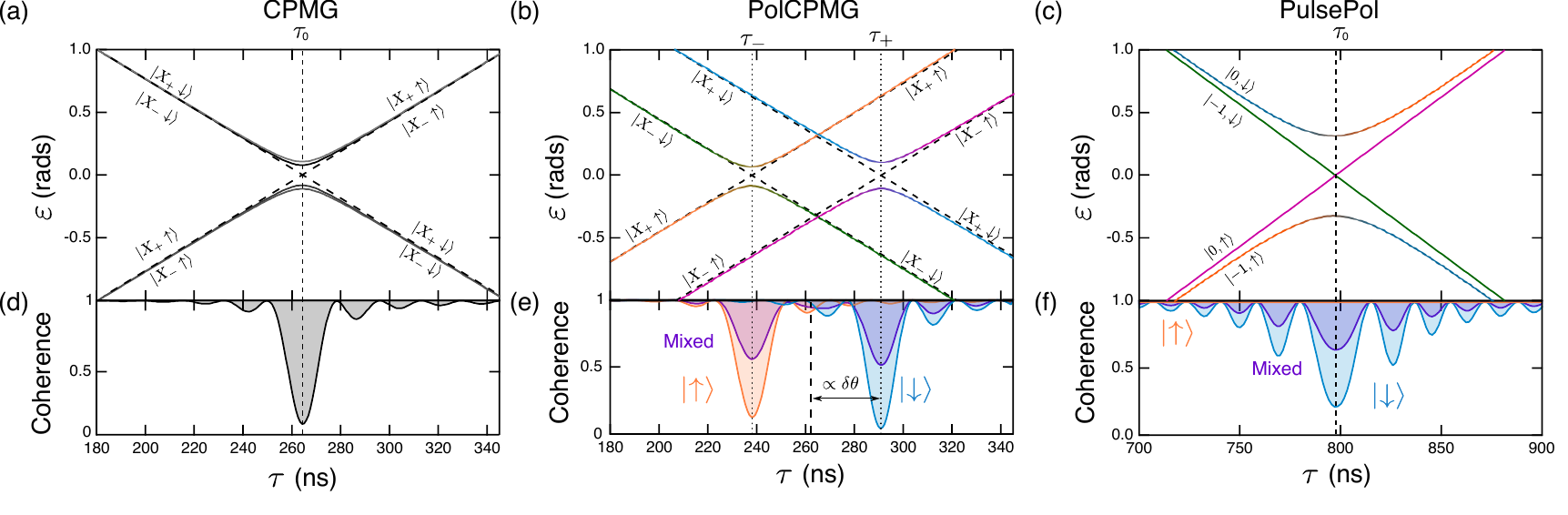}
	\caption{\textbf{Floquet spectra with finite pulse duration.} 
		(a-c) Floquet spectra of the NV-\C{13} coupled system driven by (a) CPMG without adjustment, (b) PolCPMG with $\delta\theta=\pi/10$ and (c) PulsePol. Parameters are $\omega_L/2\pi=1.9$~MHz, ${\cal A}_\perp/2\pi=180$~kHz, $t_p^0=40$~ns. The adjustment $\delta\theta$ in (b) is achieved by setting the pulse duration to $t_p=44$~ns. Dashed lines indicate the uncoupled case ($\mathcal{A}_\perp=0$).
		(d-f) Coherence of the electron spin after a CPMG (d) or PolCPMG (e) sequence comprising $N=32$ pulses, or after a PulsePol sequence comprising 12 repetitions of the unit sequence. The electron spin is initialised in \ket{X_+} in (d,e) and in \ket{0} in (f), and the plotted signal is $\langle 2\hat{S}_x \rangle$ in (d,e) whilst it is $\langle 2\hat{S}_z \rangle$ in (f). The nuclear spin is initialised in \ket{\uparrow} (orange), \ket{\downarrow} (blue) or in a completely mixed state (purple). In (d) all the different cases are overlapped and shown in black. 
	}  
	\label{Fig: Th_with_finite_tp}
\end{figure*}

\subsection{Resonance positions}
\label{sec: res pos}

The resonance conditions of PolCPMG are determined by the positions of avoided crossings in the Floquet spectrum which are in turn given by the positions of level crossings in the unperturbed Floquet spectrum. The unperturbed Floquet spectrum is given by the eigenphases of the uncoupled one-period evolution operator, $\hat{U}_{\vec{\mathcal{A}} = 0}(t = 2\tau)$. Removing the coupling term from Eq.~\eqref{eq: Ham} makes the Hamiltonian separable and the propagator is given by 
\begin{equation}
\hat{U}_{\vec{\mathcal{A}} = 0}(2\tau)= \exp(-i\omega_L\hat{I}_z2\tau)\hat{U}_p(2\tau),   \label{eq: U0 general}
\end{equation}
where $\hat{U}_p(t)$ is the pulse propagator which describes the effect of the control pulse sequence on the electronic spin (in isolation).

For top-hat pulses the one-period pulse propagator can be constructed (see Fig.~\ref{fig: PolCPMG schematic}) as follows
\begin{equation}
\hat{U}_p(2\tau) = \left[\exp\left(-i\Delta\omega\hat{S}_z\frac{\tau'}{2}\right)\exp\left(-i(\Delta\omega\hat{S}_z + \Omega\hat{S}_x)t_p\right)\exp\left(-i\Delta\omega\hat{S}_z\frac{\tau'}{2}\right)\right]^2.
\end{equation}
where $\tau' = \tau - t_p$. We define $\Omega_{\Delta\omega} = \sqrt{\Omega^2 + \Delta\omega^2}$ and $\theta_{\Delta\omega} = \arctan(\Delta\omega/\Omega)$ and compute the matrix products to find
\begin{align}
\hat{U}_p(2\tau) &= a_0\hat{\mathbb{I}} - i(a_x\hat{\sigma}_x + a_z\hat{\sigma}_z), \\
a_0 &=  \cos\frac{\Delta\omega\tau'}{2}\cos\frac{\Omega_{\Delta\omega} t_p}{2} - \sin\frac{\Delta\omega\tau'}{2}\sin\frac{\Omega_{\Delta\omega} t_p}{2}\sin \theta_{\Delta\omega} ,\\
a_x &=  \sin\frac{\Omega_{\Delta\omega} t_p}{2} \cos\theta_{\Delta\omega}, \\
a_z &=  \sin\frac{\Delta\omega\tau'}{2}\cos\frac{\Omega_{\Delta\omega} t_p}{2} + \cos\frac{\Delta\omega\tau'}{2}\sin\frac{\Omega_{\Delta\omega} t_p}{2}\sin \theta_{\Delta\omega}, \\
\end{align}
which can then be written in the form
\begin{equation}
\hat{U}_p(2\tau) = \exp(-i\delta\varepsilon_p(\tau)(\cos\theta_p\hat{S}_x + \sin\theta_p\hat{S}_z)),    \label{eq: Up general}
\end{equation} by defining
\begin{align}
\delta\varepsilon_p &= 4\arccos (a_0),  \\
\theta_p &= \arctan(\frac{a_z}{a_x}).
\end{align}

When $\Delta\omega = 0$, we get $\theta_p = 0$ and $\delta\varepsilon_p = 2\Omega t_p = 2\pi + 2\delta\theta$ where $\delta\theta$ is given by Eq.~(\ref{eq:deltatheta}). If $\delta\theta = 0$ then the sequence becomes standard CPMG and $\hat{U}_p(2\tau) = -\hat{\mathbb{I}}$. If $\delta\theta \neq 0$ then $\hat{U}_p(2\tau) = -\exp(-i2\delta\theta\hat{S}_x)$ resulting in a small $x$-rotation. The factor of 2 is present because the basic pulse unit includes two pulses. For small detunings $\Delta\omega/\Omega \ll 1$ the rotation axis is tilted slightly from the $x$-axis ($\theta_p \ll 1$).

This $x$-rotation has been well understood in the context of the error-robustnesss of CPMG \cite{Wang2012, Lang2019} and was, in fact, the motivation for designing more robust pulse sequences \cite{maudsley1986modified, gullion1990new} that suppress the accumulation of pulse errors. However, this $x$-rotation does not affect the electronic spin if it is initialised to the $\ket{X_\pm}$ states and in fact CPMG has been shown to outperform more robust sequences in protecting the coherence of these states  \cite{Ryan2010,Souza2011,Wang2012,Shim2012,Ali2013,Farfurnik2015}. For PolCPMG we exploit the effect additional pulse rotations have on the resonant electronic-nuclear spin dynamics and show how it results in polarisation transfer between the electronic and nuclear spin.

Combining Eq.~\eqref{eq: U0 general} and Eq.~\eqref{eq: Up general} we find
\begin{equation}
\hat{U}_{\vec{\mathcal{A}} = 0}(2\tau) = -\exp(-i(\omega_L\hat{I}_z 2\tau + \delta\varepsilon_p(\tau)(\cos\theta_p\hat{S}_x + \sin\theta_p\hat{S}_z))) .
\label{eq: Uunperturbed}
\end{equation}
The unperturbed Floquet spectrum is thus given by 
\begin{equation}
\varepsilon = \pm \omega_L\tau \pm \delta\varepsilon_p(\tau)/2.
\end{equation}
The eigenstates (Floquet modes) of $\hat{U}_{\vec{\mathcal{A}} = 0}(2\tau)$ are $\ket{\tilde{X}_\pm \uparrow}$ and $\ket{\tilde{X}_\pm \downarrow}$ where 
\begin{align}
\ket{\tilde{X}_+} &= \cos\left(\frac{1}{2}\left(\frac{\pi}{2} - \frac{\theta_p}{2}\right)\right)\ket{u} + \sin\left(\frac{1}{2}\left(\frac{\pi}{2} - \frac{\theta_p}{2}\right)\right)\ket{d} \approx \ket{X_+}, \\ 
\ket{\tilde{X}_-} &= \sin\left(\frac{1}{2}\left(\frac{\pi}{2} - \frac{\theta_p}{2}\right)\right)\ket{u} - \cos\left(\frac{1}{2}\left(\frac{\pi}{2} - \frac{\theta_p}{2}\right)\right)\ket{d} \approx \ket{X_-}
\end{align}
where the approximations are exact for $\Delta\omega = 0$.

At $\tau = \tau_+$ there is a level crossing between the $\ket{\tilde{X}_+\downarrow}$ and $\ket{\tilde{X}_-\uparrow}$ Floquet modes. At $\tau = \tau_-$ there is a level crossing between the $\ket{\tilde{X}_+\uparrow}$ and $\ket{\tilde{X}_-\downarrow}$ Floquet modes. To find the dip positions we find where the Floquet phases are equal by solving
\begin{equation}
\mp\omega_L\tau_\pm + \delta\varepsilon_p(\tau_\pm)/2 = \pm\omega_L\tau_\pm - \delta\varepsilon_p(\tau_\pm)/2.
\label{eq: taupm}
\end{equation}
Typically $\delta\varepsilon_p$ will be some complex analytic function of $\tau$ (see below for the top-hat pulse case) but Eq.~\eqref{eq: taupm} can be solved numerically. When $\Delta\omega = 0$, we have $\delta\varepsilon_p = 2\Omega t_p$ and the dip positions can be obtained analytically, 
\begin{equation}
\tau_\pm = \frac{\pi \pm \delta\theta}{\omega_L} = \tau_0\left(1 + \frac{\delta\theta}{\pi}  \right).
\label{eq: taupm2}
\end{equation}
In the standard CPMG case $\delta\theta = 0$, we recover the expected single dip position. Fig.~\ref{fig: res pos} plots the expected dip positions for a range of pulse widths and detunings. When $\Delta\omega \neq 0$ (Fig.~\ref{fig: res pos}b) the expected dip positions are plotted by numerically solving Eq.~\eqref{eq: taupm}. The analytic prediction for the resonance positions shows a good fit with numerics. 

\begin{figure*}[h!]
\includegraphics[scale = 0.8]{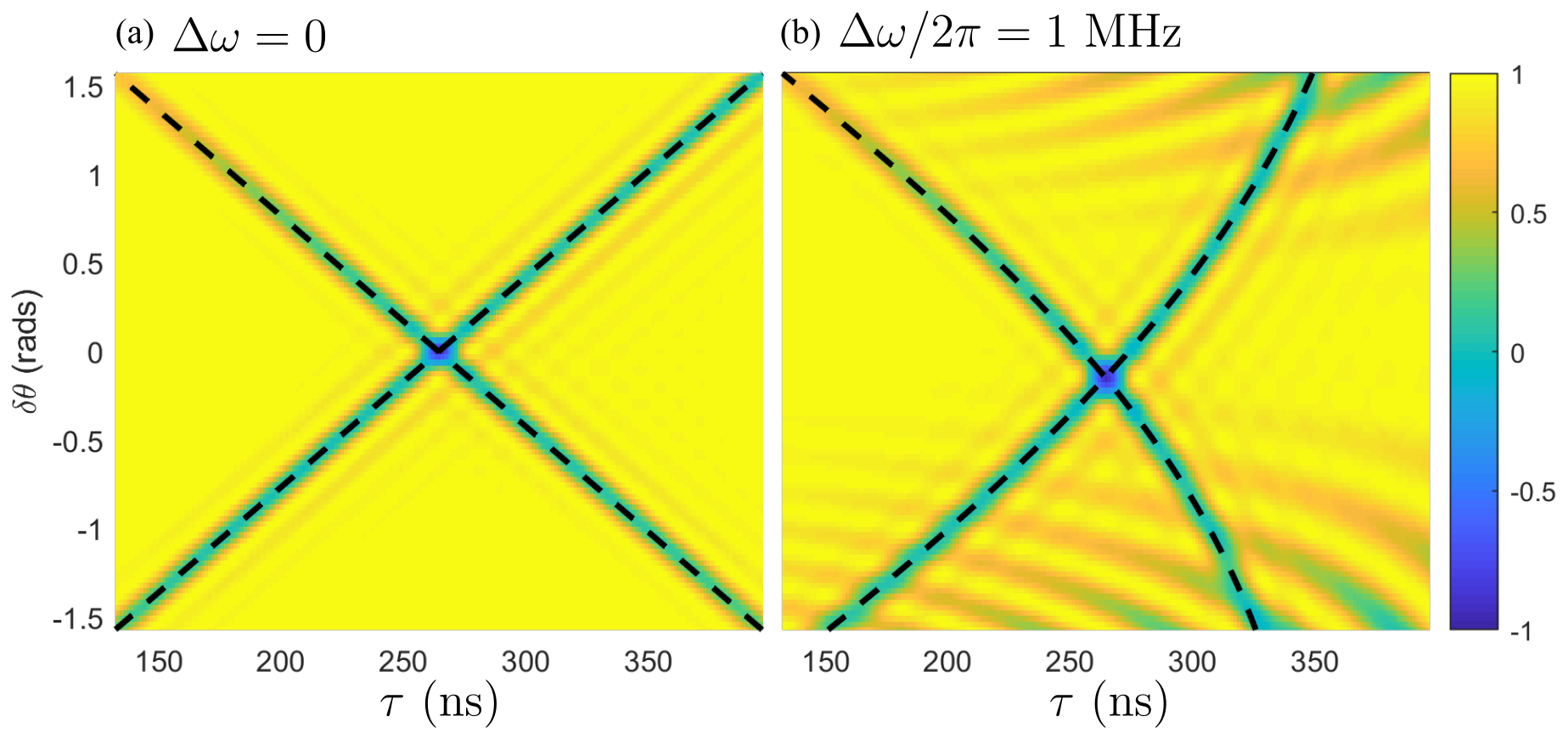}
\caption{\textbf{PolCPMG coherence maps.} Numerical simulation of the NV coherence after PolCPMG for a scan of rotation adjustments $\delta\theta$ as defined by Eq.~(\ref{eq:deltatheta}). In (a) there is no detuning ($\Delta\omega = 0$) and the superimposed black dashed lines show the dip position calculated via Eq.~\eqref{eq: taupm2}. In (b) $\Delta\omega/2\pi = 1$~MHz and the dip positions are calculated by numerically solving Eq.~\eqref{eq: taupm}. For simulation here we model an NV coupled to a \C{13} with $A_\perp/2\pi = 180$~kHz, $\omega_L/2\pi = 1.9$~MHz, $t_p^0 = 40$~ns and $N = 32$.}  
\label{fig: res pos}
\end{figure*}

The benefit of using Floquet theory is that once we have analytically determined the unperturbed Floquet spectrum and states it is clear from numeric plots of the full Floquet spectrum (e.g. Fig.~\ref{Fig: Th_with_finite_tp}(b)) what the coherence effect will be. At the start of the PolCPMG protocol the electronic spin is initialised into the $\ket{X_+}$ state but the nuclear spin is in a thermal mixture of $\ket{\uparrow}$ and $\ket{\downarrow}$. The joint system is thus in a thermal mixture of $\ket{X_+\uparrow}$ and $\ket{X_+\downarrow}$.

At $\tau = \tau_+$, say, the $\ket{X_+ \uparrow}$ portion of the initial mixture is in an eigenstate of the PolCPMG control but the $\ket{X_+ \downarrow}$ portion couples to the $\ket{X_- \uparrow}$ state as indicated by the avoided crossing. The initial state $\ket{X_+\uparrow}//\ket{X_+ \downarrow}$ thus evolves under PolCPMG to $\ket{X_+\uparrow}// (\cos(r_+N_p)\ket{X_+\downarrow} + \sin(r_+N_p)\ket{X_-\uparrow})$ after $N_p$ repetitions and where $r_+$ is the polarisation rate determined in the next section. The coherence ($\mathcal{L} = \langle 2\hat{S}_x \rangle$) and nuclear polarisation ($\mathcal{P} = \langle 2\hat{I}_z \rangle$) of this state are 
\begin{align}
\mathcal{L} &= 1 - \sin^2 \left(r_+N_p\right), \\
\mathcal{P} &= \sin^2 \left(r_+N_p\right).   \label{eq: P}
\end{align}
The nuclear spin is thus fully polarised after $N = 2N_p = \pi/r_+$ pulses and this happens at time $t_\text{pol}^+ = N\tau_+ = \pi\tau_+/r_+$.

At $\tau_-$ the nuclear spin will be polarised into the opposite state. In both cases the polarisation time is given by 
\begin{equation}
t^\pm_\text{pol} = \frac{\pi\tau_\pm}{r_\pm},
\end{equation}
where $r_\pm$ is the polarisation rate, to be determined in the next section. If we have $\ket{X_\pm} \neq \ket{\tilde{X}_\pm}$, this will affect the fidelity of a single PolCPMG sequence but not the polarisation rate. By reinitialising the electronic spin and repeating the sequence (within the $T_1$ time of the nuclear spin) the fidelity can be restored. 

Note that running the PolCPMG protocol with $\tau \neq \tau_\pm$ we see that both portions of the initial mixed state are eigenstates of the PolCPMG protocol and the full initial state is protected -- thus the PolCPMG sequence can also function as dynamical decoupling by simply changing the pulse spacing.

\subsection{Polarisation rate}
\label{sec: pol rate}

To find the polarisation rate we need to determine the full ($\vec{\mathcal{A}} \neq 0$) one-period propagator at the resonance positions, $\tau_\pm$. This propagator can be found by concatenating sections of free evolution, $\hat{U}_0(t) = \exp(-i(\omega_\text{L}\hat{I}_z + \hat{S}_z\vec{\mathcal{A}}\cdot\vec{I})t)$ with the pulse operator $\hat{U}_{\tilde{\pi}} = \exp(-i\delta\varepsilon_p/2(\cos\theta_p\hat{S}_x + \sin\theta_p\hat{S}_z))$ in the order $\hat{U}(2\tau) = [\hat{U}_0(\tau/2)\hat{U}_{\tilde{\pi}}\hat{U}_0(\tau/2)]^2$ (see Fig.~\ref{fig: PolCPMG schematic}). Here we have assumed that the pulses are instantaneous but the additional rotation $\delta\theta$ and detuning $\Delta\omega$ are still present. 

At $\tau = \tau_+$ say, we know from the Floquet spectrum that the dynamics can be reduced to a 2D pseudospin-1/2 in the $\{\ket{\tilde{X}_+\downarrow}, \ket{\tilde{X}_-\uparrow}\}$ subspace. In this subspace the evolution operators are given by 
\begin{align}
\hat{U}_0(t) &\approx 
\begin{bmatrix}
e^{+i\omega_L\tau/4} & 0 \\
0 & e^{-i\omega_L\tau/4}
\end{bmatrix}
-i\frac{A_\perp}{\omega_L}
\begin{bmatrix}
0 & g \\
g & 0
\end{bmatrix} 
+ \mathcal{O}\{\left(\frac{A_\perp}{\omega_L}\right)^2\}
\\
\hat{U}_{\tilde{\pi}} &= 
\begin{bmatrix}
e^{-i\delta\varepsilon/4} & 0 \\
0 & e^{+i\delta\varepsilon/4}
\end{bmatrix}
\end{align}
where $g = \frac{1}{2}\cos\theta_p\sin(\omega_L \tau/4)$ and we ignore the contribution from $A_\parallel\hat{I}_z$ as it is cancelled up to higher order by the PolCPMG sequence. Concatenating these sections of evolution in the order specified previously we find
\begin{equation}
\hat{U}(2\tau) = 
\begin{bmatrix}
e^{-i(\delta\varepsilon/2 - \omega_L\tau)} & 0 \\
0 & e^{+i(\delta\varepsilon/2 - \omega_L\tau)} \\
\end{bmatrix}
-i\frac{A_\perp}{\omega_L}
\begin{bmatrix}
0 & G \\
G & 0 
\end{bmatrix}
+ \mathcal{O}\{\left(\frac{A_\perp}{\omega_L}\right)^2\}
\end{equation}
where $G = \cos\theta_p\sin(\omega_L \tau/4) (\cos(\omega_L \tau/4) + \cos((2 \delta\varepsilon_p - 3 \omega_L \tau)/4))$. This full ($\vec{\mathcal{A}} \neq 0 $) one-period propagator,  is equivalent to the uncoupled ($\vec{\mathcal{A}} = 0$) one-period propagator, Eq.~\eqref{eq: Uunperturbed}) but with an additional coupling term. This additional coupling is what creates the avoided crossings in the Floquet spectra and it induces the polarisation transfer. 

At $\tau = \tau_+$, say, the initial state $\ket{\tilde{X}_+ \downarrow}$ evolves under one repetition of PolCPMG to $\ket{\tilde{X}_+\downarrow} - i \frac{A_\perp}{\omega_L}G \ket{\tilde{X}_-\uparrow} \approx \cos(\frac{A_\perp}{\omega_L}G)\ket{\tilde{X}_+\downarrow} - i \sin(\frac{A_\perp}{\omega_L}G) \ket{\tilde{X}_-\uparrow}$. Thus, by comparison with the discussion of $r_\pm$ in the previous section we find that 
\begin{equation}
r_\pm = \frac{A_\perp}{\omega_L}G_\pm. \label{eq: polrate} 
\end{equation}
where $G_\pm = G(\tau_\pm)$. When $\Delta\omega = 0$, $G_\pm = \cos\frac{\delta\theta}{2}$ and the polarisation time along each branch is thus 
\begin{equation}
t_\text{pol}^\pm = \frac{\pi\tau_\pm}{r_\pm} = \frac{\pi}{A_\perp\cos\frac{\delta\theta}{2}}\left(\pi \pm \delta\theta \right) \approx \frac{\pi^2}{A_\perp}.
\label{eq: tpol}
\end{equation}
The expression for the polarisation rate is independent of the particular dip branch (i.e. $\tau_+$ and $\tau_-$ have the same polarisation rate) however as $\tau_- < \tau_+$ the total polarisation time on this branch will be shorter as indicated in Eq.~\eqref{eq: tpol}. Numerical simulations show that further asymmetry can occur between the branches when finite-pulse-durations and detunings are taken into account. Whilst this fine tunes the polarisation rate of PolCPMG the basic underlying physical principle remains the same. Note that the $A_\parallel$ term in $\vec{\mathcal{A}} = (A_\perp, 0, A_\parallel)$ only contributes to the dynamics at higher order so only $A_\perp$ is present in our expression for the polarisation rate.  

PolCPMG has comparable polarisation time to NOVEL \cite{London2013} ($t_\text{pol} = 2\pi/A_\perp$) and PulsePol \cite{Schwartz2018} ($t_\text{pol} = 2\pi/(\alpha A_\perp)$, $\alpha = \frac{2}{3\pi}(2 + \sqrt{2})\approx0.72$) with an improved robustness to detunings over NOVEL and an improved robustness to phase errors over PulsePol. 

Figure~\ref{fig: Pol Analytic} compares numerical simulations of the nuclear polarisation under PolCPMG with a conjunction of the analytic predictions presented here -- namely Eqs.~\eqref{eq: P}, \eqref{eq: polrate} and \eqref{eq: tpol}. The analytics show an excellent fit to numerics. The simulations in the main text include inhomogeneous broadening, extra detunings due to NV's host nitrogen spin (see details in Sec.~\ref{sec:ExpDriving}) and finite-duration pulse widths which all serve to perturb the polarisation dynamics. However, the basic underlying principle remains the same. 

\begin{figure*}[h]
\includegraphics[scale = 0.7]{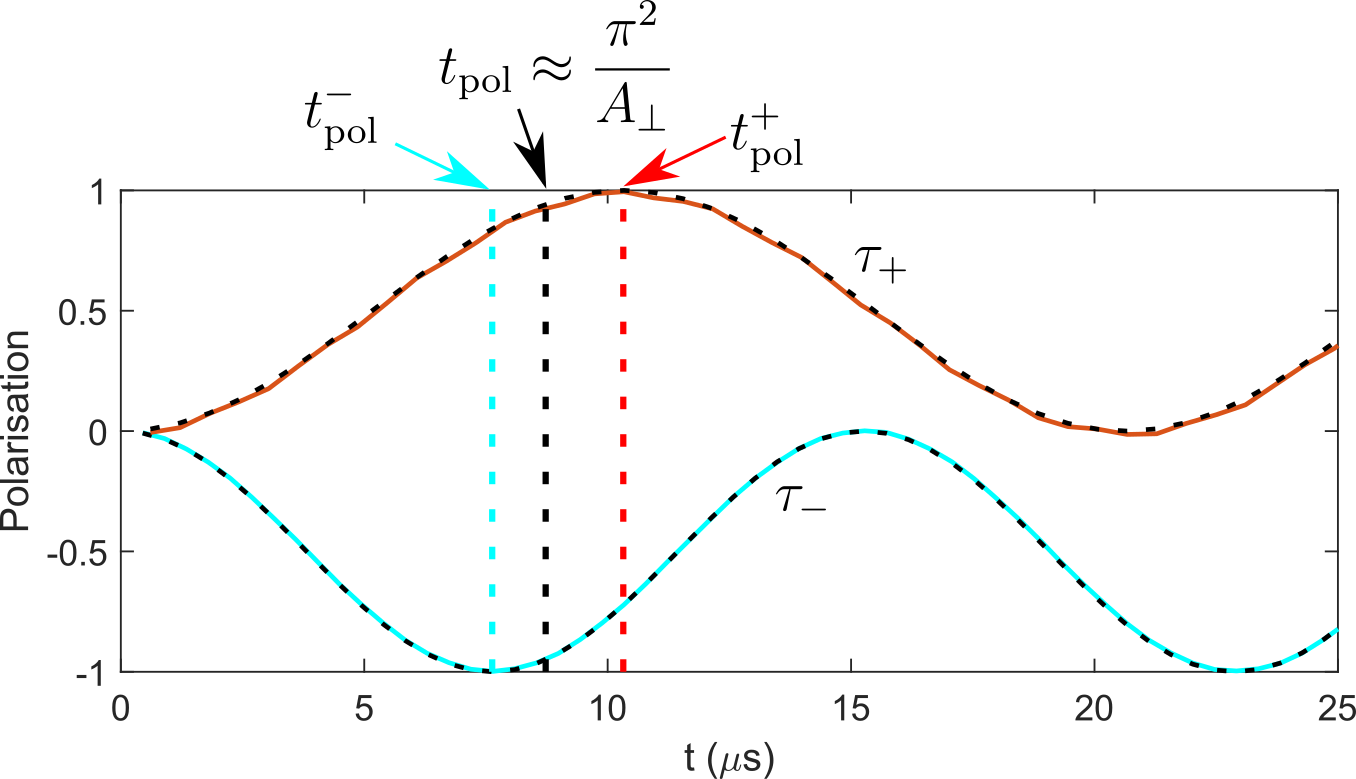}
\caption{\textbf{Nuclear polarisation under PolCPMG.} Graph showing the nuclear polarisation after PolCPMG applied at $\tau_+$ (red solid line) and $\tau_-$ (blue solid line). Here the number of pulses is scanned and plotted as $t = N\tau$. The solid lines are calculated via numeric simulation of PolCPMG at $\delta\theta = 0.15\pi$ with $\tau_\pm$ determined via Eq.~\eqref{eq: taupm2}. The black dashed lines represent the analytic predictions from Eq.~\eqref{eq: P}. The vertical dashed lines indicate $t_\text{pol}^\pm$ and the indicative polarisation time $t_\text{pol} = \pi^2/A_\perp$ (see Eq.~\eqref{eq: tpol}). We simulate an NV coupled to a $^{13}$C spin via $A_\perp/2\pi = 180$~kHz at $\omega_L = 1.9$~MHz.}  
\label{fig: Pol Analytic}
\end{figure*}

\subsection{Robustness to frequency detunings}

In Fig. 4 of the main text, we presented simulations of the nuclear polarisation as a function of three types of control errors: (i) frequency detuning $\Delta\omega$, (ii) power detuning $\Delta\Omega=\Omega-\Omega_0$ where $\Omega$ is the microwave driving strength and $\Omega_0$ is the actual electronic Rabi frequency when $\Delta\omega=0$, and (iii) rotation axis error (or phase error) $\Delta\phi=(\phi_y-\phi_x)-\pi/2$ where $\phi_x$ ($\phi_y$) is the microwave phase during the $x$-pulses ($y$-pulses).

The robustness to frequency detunings deserves special discussion as it depends on the choice of the pulse duration adjustment $\delta\theta$. To illustrate this, Fig.~\ref{fig: OWP} simulates sensor coherence maps after PolCPMG -- scanning pulse spacing $\tau$ and the microwave detuning $\Delta\omega$ -- for three different pulse rotation adjustments $\delta\theta = 0, +22.5^\circ, - 22.5^\circ$. The resonance positions $\tau_\pm \equiv \tau_\pm(\delta\theta, \Delta\omega)$ depend sensitively on the additional rotation and detuning. However, for $\delta\theta \approx +22.5^\circ$ it can be seen that the resonance positions are markedly less sensitive to detuning errors about $\Delta\omega = 0$. This insensitivity is particularly interesting in the presence of inhomogeneous broadening (and an unpolarised host nitrogen in NV experiments) where the signal is averaged over a range of detuning values. Selecting $\delta\theta \approx +22.5^\circ$ provides an optimal working point where the system is more robust to detuning errors.

This effect can be clearly seen in Fig.~3d in the main text where the nuclear polarisation is measured at $\tau_\pm(\delta\theta)$ for a scan of $\delta\theta$. The peak in polarisation on each branch at about $\delta\theta \approx +22.5^\circ$ is due to the insensitivity to $\Delta\omega$. In Fig.~3d in the main text both experiment and simulation included unpolarised $^{14}$N and inhomogeneous broadening.

\begin{figure*}[h!]
\includegraphics[scale = 0.9]{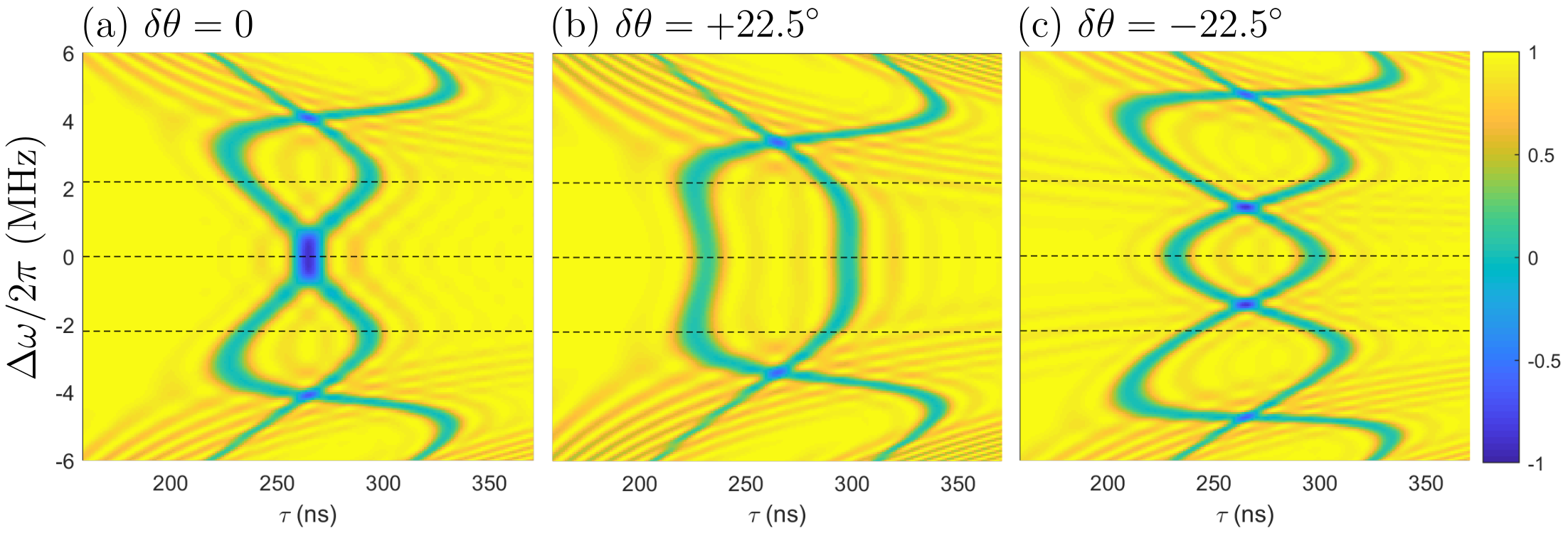}
\caption{\textbf{PolCPMG optimal working points.} Numerical simulation of the NV coherence after PolCPMG for a scan of detuning errors $\Delta\omega$. In (a) there is no additional rotation, $\delta\theta = 0$. In (b) $\delta\theta = +22.5^\circ$ and the resonance positions are shown to be more stable about $\Delta\omega = 0$. In (c) $\delta\theta = -22.5^\circ$ and the resonance positions are more sensitive to small changes in $\Delta\omega$. The dashed black lines represent the detunings provided by the unpolarised $^{14}$N. For simulation here we model an NV coupled to a \C{13} with $A_\perp/2\pi = 180$~kHz, $\omega_L/2\pi = 1.9$~MHz, $t_p^0 = 40$~ns and $N = 32$. In (a) $t_p = 40$~ns, (b) $t_p = 45$~ns, (c) $t_p = 35$~ns.}  
\label{fig: OWP}
\end{figure*}

\newpage

\section{Experimental realisation of PolCPMG}

\subsection{Diamond sample}

The NV-diamond sample used in these experiments was a 2~mm $\times$ 2~mm $\times$ 50 $\mu$m electronic grade single-crystal diamond with $\left\lbrace 110\right\rbrace$ edges and a (001) top facet. The NV spins were introduced through an overgrowth with gas flow rates of 1000~sccm H$_2$, 20~sccm CH$_4$, 10~sccm N$_2$, at a pressure of 100~Torr, with a microwave power of 3000~W producing a sample temperature of $T \approx 900~^\circ$C. The total time of growth was 6~mins at a rate of $10~\mu$m/hr thus introducing a 1~$\mu$m-thick layer of ensemble NV spins. The overgrown diamond had a natural isotopic abundance, $[^{13}$C$]=1.1\%$.  

\subsection{Experimental apparatus}

The experiments were conducted at room temperature with a custom-built wide-field fluorescence microscope described in Refs.~\cite{Broadway2018b, Broadway2018c}. The diamond was glued to a glass cover slip with patterned microwave waveguide. Optical excitation was performed with a 532~nm Verdi laser that was gated using an acousto-optic modulator (AA Opto-Electronic MQ180-A0,25-VIS) and focused to the back aperture of an oil immersion objective lens (Nikon CFI S Fluor 40x, NA = 1.3). The photoluminescence (PL) from the NV centres is separated from the excitation light with a dichroic mirror and filtered using a bandpass filter before being imaged using a tube lens (f = 300 mm) onto a sCMOS camera (Andor Zyla 5.5-W USB3). Microwave excitation was provided by a signal generator (Rohde \& Schwarz SMBV100A) gated using the built-in IQ modulation and amplified (Mini-circuits HPA-50W-63+). A pulse pattern generator (Spincore PulseBlasterESR-PRO 500 MHz) was used to gate the excitation laser, microwaves, and to synchronise the image acquisition. A static magnetic field of strength $B = 1765$~G was applied along the NV axis using a permanent magnet. 

In all experiments, the laser spot diameter was about $100~\mu$m at the NV layer and the total CW laser power at the sample was 300 mW. The laser pulse duration for each NV spin initialisation/readout was $20~\mu$s, chosen as a trade-off between readout contrast and initialisation fidelity. In main text Figs. 2 and 3, the PL signal was averaged over a small region ($1.6~\mu$m~$\times 1.6~\mu$m) to avoid issues arising from spatial inhomogeneities (especially in microwave frequency and power). Such an area corresponds to an estimated total of $10^5$ NV centres. The exposure time of the camera was 1 ms and the total acquisition time tens of minutes to hours for each measurement. In main text Fig. 4, a $30~\mu$m~$\times 30~\mu$m area was imaged and analysed. Each $0.4~\mu$m~$\times 0.4~\mu$m pixel in the image contained about $10^4$ NV centres. The exposure time of the camera was 10 ms and the total acquisition time about ten hours.

\subsection{Implementation of polarisation measurements}

For each experiment presented in the paper, the overall sequence was adapted to accommodate the time scale mismatch between the duration of a single polarisation cycle (initialise-PolCPMG-readout) and the exposure time of the camera (Fig.~\ref{Fig: Sequence}a). The building blocks of any sequence consisted of $R$ repeats of the initialise-PolCPMG-readout sequence where the NV spins are initialised in the \ket{X_+} state (Fig.~\ref{Fig: Sequence}b) or in the \ket{X_-} state (Fig.~\ref{Fig: Sequence}c). To measure the NV coherence while preventing \C{13} polarisation build-up as in main text Fig. 2c (top panel), the two initialised states are simply alternated during the camera exposure (Fig.~\ref{Fig: Sequence}d). For all the other experiments, we first reset the \C{13} bath by applying $R=500$ cycles with \ket{X_-} (or alternating \ket{X_+} and \ket{X_-}) before applying the sequence depicted in main text Fig. 2b, i.e. we apply $R$ cycles of a test sequence with \ket{X_+} and variable parameters ($t_p$, $\tau$, $N$ etc.) and then perform a single \ket{X_+} sequence during camera exposure to readout the \C{13} polarisation state (Fig.~\ref{Fig: Sequence}e). In all cases, the signal is normalised by repeating the same measurement but changing only the final projecting pulse from a $\pi/2$ to a $3\pi/2$, allowing us to infer the probability of finding the NV in the \ket{0} state, which is a measure of the NV coherence before this final projecting pulse.

\begin{figure*}[h]
	\includegraphics{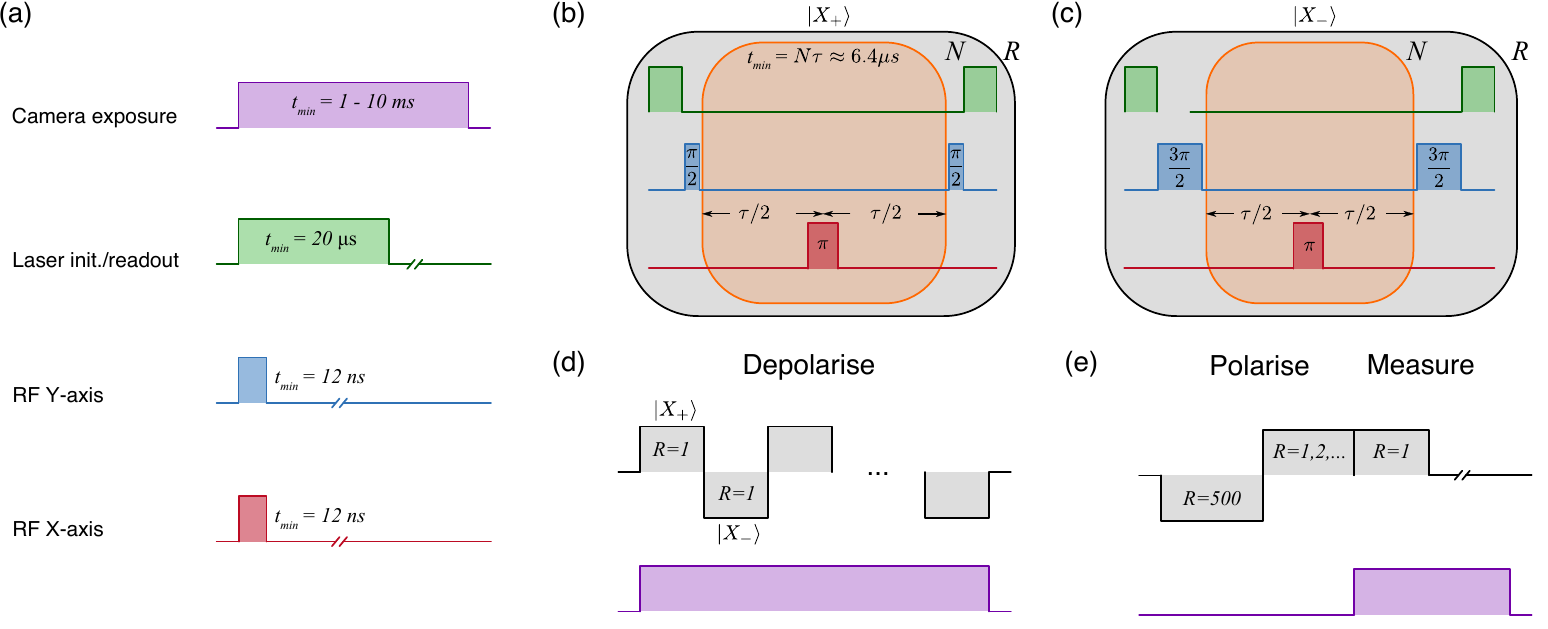}
	\caption{\textbf{Hyperpolarisation sequences with a camera.} 
		(a) Legend of pulse sequence including with minimum time for each action.
		(b) PolCPMG sequence, which comprises a laser initialisation, a $\pi/2$ rotation to initialise in \ket{X_+}, a series of $\pi$-rotations (where the rotation adjustments are introduced), a second $\pi/2$ rotation to project into the $z$-basis and a final readout laser pulse.
		(c) As in (b) except with an initial $3\pi/2$ rotation to initialise in \ket{X_-}, producing the opposite polarisation.  
		(d) Depolarisation sequence, in which initialisation is alternated between \ket{X_+} and \ket{X_-} to prevent polarisation build-up.
		(e) Polarisation sequence, in which the bath is initially polarised in one direction (no camera exposure) then the polarisation build-up with the opposite initialisation is measured. The individual components have independent parameters ($N, \tau, R$), facilitating the parameter exploration in the main text.  
	}  
	\label{Fig: Sequence}
\end{figure*}

\subsection{Driving of the NV spin} \label{sec:ExpDriving}

The \ket{0}$\rightleftharpoons$\ket{-1} manifold of the NV electron spin resonance has a frequency centred around $\omega_{\rm NV}=|D-\gamma_e B|$ where $D=2870$~MHz is the zero-field splitting and $\gamma_e=2.8$~MHz/G the electron gyromagnetic ratio. In the main text, a magnet field of $B = 1765$~G is applied resulting in $\omega \approx 2078$~MHz. In addition, The NV spin exhibits a hyperfine structure due to the $^{14}$N nuclear spin (spin-1) with a hyperfine constant $A_{^{14}{\rm N}} = 2.2$~MHz. That is, in general there are three electron spin resonance frequencies: $\omega_{\rm NV}$, $\omega_{\rm NV}\pm A_{^{14}{\rm N}}$. When operating in regimes where the nuclear spin is not optically polarised~\cite{Broadway2016,Ivady2015,Jacques2009}, the presence of this hyperfine structure impacts the response of the PolCPMG sequence and needs to be considered. Prior to the experiments, an electron spin resonance spectrum was recorded at low microwave power to identify the central frequency $\omega_{\rm NV}$ (Fig.~\ref{Fig: Hyperfine Pol}). In the subsequent PolCPMG experiments, the microwave driving frequency is set to $\omega_{\rm NV}$ and the microwave power is adjusted so as to obtain a Rabi frequency $\Omega=12.5$~MHz~$\gg A_{^{14}{\rm N}}$. The spectrum in Fig.~\ref{Fig: Hyperfine Pol} reveals a partial polarisation of the $^{14}$N nuclear spin generated due to state mixing from the distant ground state level anti-crossing. By fitting the spectrum we find populations (0.5, 0.3, 0.2) for the \ket{-1,0,+1} hyperfine states, respectively, and a broadening of each line of about 1~MHz (FWHM). These values were used in the simulations in main text Figs. 2d, 3c and 3d.

\begin{figure*}[h]
	\includegraphics{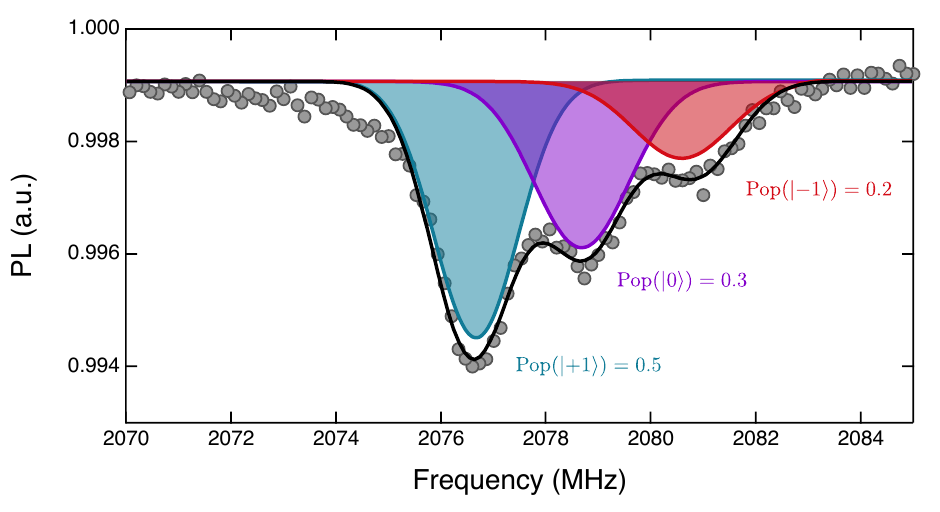}
	\caption{\textbf{NV hyperfine polarisation.}
		High-resolution electron spin resonance spectrum taken at $B = 1765$~G indicating a residual polarisation of the $^{14}$N nuclear spin from the GSLAC. The populations for each $^{14}$N state ($m_I=0,\pm1$) are displayed. The driving frequency ($\omega$) for PolCPMG experiments is indicated with a dashed line. 
	}
	\label{Fig: Hyperfine Pol}
\end{figure*}

\subsection{Introduction of spatially varying frequency and power detuning }

In main text Fig. 4c, spatial gradients in NV frequency and microwave power were deliberately applied to the sample and an image was taken over a $30~\mu$m~$\times 30~\mu$m region (Fig.~\ref{Fig: Maps}a). The background magnetic field (generated by a permanent magnet) needs to be aligned with the NV spin $z$-axis which is 54.7$^\circ$ from the surface normal, which introduces a gradient in the NV frequency (Fig.~\ref{Fig: Maps}b). The power detuning was generated by applying the microwave driving through a strip line, causing a $1/r$ decay in the driving power (Fig.~\ref{Fig: Maps}c). The combination of these two detunings produce the polarisation map (Fig.~4 main text and Fig.~\ref{Fig: Maps}d). We note that for larger scale applications of hyperpolarisation, these gradients can be minimised by using electromagnets instead of permanent magnets, and properly designed microwave antennas instead of a strip line~\cite{Herrmann2016}. 

\begin{figure*}[h]
	\includegraphics{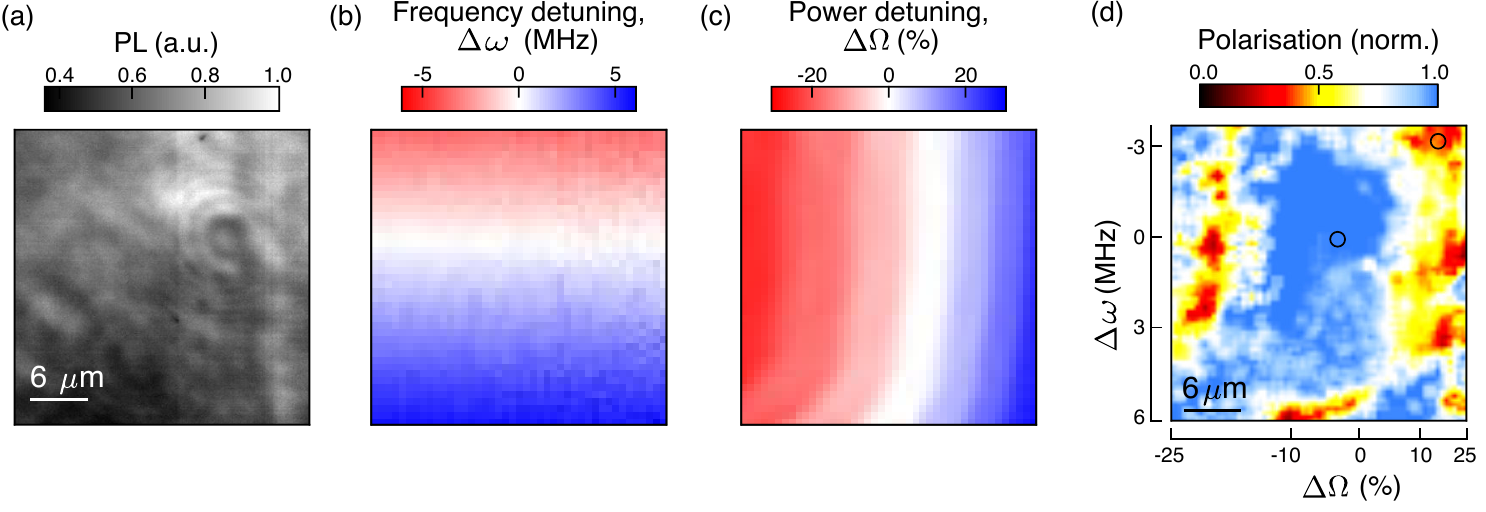}
	\caption{\textbf{Spatial maps of detuning.} 
	(a) PL map of the $30\times30~\mu$m region imaged in main text Fig. 4c. 
	(b) Map of the frequency detuning ($\Delta\omega$) across the region of interest, i.e. the difference between the NV frequency $\omega_{\rm NV}$ and the microwave driving frequency ($\omega = 2078.2$~MHz) for PolCPMG. 
	(c) Variation in Rabi frequency ($\Delta\Omega$) relative to the desired driving strength of $\Omega = 12.5$~MHz. The gradient is generated by the decaying magnetic field strength from a wire located on the right that runs parallel to the image. Additional variation in the driving field is seen towards the bottom by frequency detuning.
	}
	\label{Fig: Maps}
\end{figure*}

\end{widetext}

\end{document}